\documentclass[pra,aps,twocolumns,superscriptaddress,showpacs,footnotes]{revtex4}

\begin{document}

\title{Bell's theorem tells us {\em not} what quantum mechanics {\em is}, but what quantum mechanics {\em is not}.}

\author{Marek \.Zukowski}
\address{Institute of Theoretical Physics and Astrophysics, University of Gda\'nsk,
ul. Wita Stwosza 57, PL-80-952 Gda\'nsk, Poland}

\begin{abstract}
Non-locality, or quantum-non-locality,  are buzzwords in the community of quantum foundation and information scientists, which purportedly describe the implications of Bell's theorem.  When such phrases are treated seriously, that is it is claimed that Bell's theorem reveals non-locality as an inherent trait of the quantum description of the micro-world, this leads to logical contradictions, which will be discussed here. In fact, Bell's theorem, understood as violation of Bell inequalities by quantum predictions, is consistent with Bohr's notion of complementarity. Thus, if it points to anything, then it is rather the significance of the principle of Bohr, but even this is not a clear implication. Non-locality is a necessary consequence of Bell's theorem only if we reject complementarity by adopting some form of realism, be it additional hidden variables, additional hidden causes, etc., or counterfactual definiteness.

The essay contains two largely independent parts. The first one is addressed to any reader interested in the topic. The second, discussing the notion of local causality, is addressed to people working in the field.
\end{abstract}


\maketitle

\section{Introduction}
In the first part I shall present a simple version of the Bell's theorem based on the works of GHZ \cite{GHZ} and Mermin \cite{MERMIN}, and its relation to the EPR paradox \cite{EPR}. Simple but not oversimplified, I hope. The aim of the presentation will be to show that the EPR work does not contain a complete analysis of the problem, and some crucial assumptions are not overtly expressed therein, but can nevertheless be pinpointed. The whole Section is formulated in as a counter-factual story about EPR attempting to write their paper, but  learning about GHZ results before finishing it. I have chosen GHZ paradox, instead of Bell's original work on "On the Einstein-Podolsky-Rosen" paradox \cite{BELL1964}, only because GHZ-type reasoning  is very accessible, and does not require any knowledge about statistical methods, or probability. 
 The presentation of the situation is for laypersons. It will require only elementary algebra, and logical intuition. The second Section  is also elementary, however it is addressed to scientists interested in Bell's theorem.

All what I present here is not new. It is a concise presentation of the basic facts about quantum theory, Einstein-Podolsky-Rosen Paradox (EPR) and Bell's Theorem. 
The sole aim is to reveal logical inconsistency in attempts to  claim that the sole assumptions of Bell's theorem are locality, freedom to choose the setting of experimental devices, and quantum predictions. Some authors following such views claim that no other assumption is needed at all (currently some claim that it is enough to assume local causality, which is correct, but they seemingly treat it  as the opposite notion to non-local causality, which not the case). Some claim the missing assumption of hidden variable or realism, or counterfactual definiteness \cite{MUYNK} is irrelevant because it can be replaced in the derivations of Bell's inequalities by determinism, which in turn is derivable via an EPR type reasoning \cite{NORSEN}.  This will be shown to be impossible in the first part of this essay.

The second section concentrates on showing that the assumption of {\em local causality} is effectively a version of local hidden variable theories. What is important, its negation is {\em not} non-local causality. As local causality is a compound condition, it may not hold because of several reasons. This may be non-locality or non-existence of causes of elementary quantum events which are outside of the quantum description (that is, that the only predictive tool describing given situation is the quantum state, and quantum formalism allowing, via some kind of form of Born rule, to calculate the probabilities). 
Thus the violation of Bell inequalities, which can be derived using freedom assumption and local causality, does not indicate whether quantum mechanics it is local or non-local.
The property of non-signalling indicates that quantum mechanics is local. While some claim the collapse of the wave packet postulate (projection postulate) points to non-locality. However, the wave function has real status only in realistic interpretations. In (neo?) Copenhagen Interpretation, which is,  admittedly, different for every apostle of it, this is just symbolic mathematical tool allowing predictions for a system which is a member of an ensemble defined by the preparation procedure. At least Bohr himself says: "actual calculations are most conveniently carried out with the help of a Schroedinger state function, from which the statistical laws governing observations obtainable under specified conditions can be deduced by definite mathematical operations. It must be recognized, however, that we are dealing here with a purely symbolic procedure, the unambiguous physical interpretation of which in the last resort requires reference to the complete experimental arrangement."\cite{BOHR1963} Note also, that the projection postulate within Qbist Interpretation is just the quantum Bayesian update \cite{QBISM}, thus cannot be treated as a non-local real phenomenon. 


\section{Paradoxes of interpretations of the EPR paradox: elementary presentation} 

Let me now present a reasoning which is based on some quantum predictions, and the assumptions of EPR which led Bell in 1964 to formulate his original inequalities, which in turn  led to his Theorem. In a way this will be an intellectual game. Let start just like EPR did in 1935, but let us consider the fascinating properties of three photons which have maximally entangled polarizations, discovered 54 years later by Greenberger, Horne and Zeilinger (GHZ). Let us see whether EPR can reach the same conclusions as in their 1935 paper, where they considered a completely different state. Let us also see whether Bohr was right in his 1935 criticism of the EPR paper \cite{BOHR1935}.  

Consider first, the physical situation described by GHZ. Imagine that we have a source,
which in a single run of the experiment, when excited, emits three particles, each in a different direction, such they reach three observers  (for a history of efforts to actually build such source with quantum optical techniques see \cite{PAN}). The three observes Danny, Mike and Anton are very very far away from each other. The photons reach their labs at the same moment of time. Spin is a kind of internal angular momentum of a particle. Angular momentum is a vector, thus spin must be a vector too. But this is a very strange vector. In the case of photons if we measure any of its components the results are always $+1$ or $-1$ (times Planck Constant dividend by $2\pi$, but this we shall treat as an irrelevant detail). By the way, polarization of light is a consequence of the fact that photons are massless "spin-1" particles, and measurements of the spin components of photons are in fact specific measurements of their polarization.  

After very many repetitions of their experiments, on a  big enough statistical sample of the emissions of  the particle-triples, the  three observes, if they inform each other about the results of their measurements and the settings of their polarization measuring local apparatuses (which is decided randomly  by the whim of each local observer separately, in each run of the experiment), notice the following phenomena. 
\begin{itemize}
\item
If two of them decide to measure the vertical component of the spin of the arriving particle (its velocity is assumed to define the third direction in space for the local observer), and remaining o third observer (whoever he is) chooses to measure the horizontal component (at the right angle with the previous one, and perpendicular to the velocity of the local particle), then the product of their local results is always $-1$. That is either all of them get $-1$ or for two of them the  results are $1$ and for one of them $-1$. 
\item
However in such situations. as well as in the other one considered below, each observer, before he exchanges the information with his partners,  sees all his $\pm1$ local results as fully random, following the statistics of fair coin toss. This is so irrespectively of the local setting of the measuring device (that is, of which component is measured).
\item
If by chance all three decide to measure the horizontal components of local spins, the product of their local results is always $1$: either all of them get $1$, or for two of them results are $-1$, and for one of them is $1$.  
\end{itemize}

We have considered four experimental situations. EPR would notice here that in each such a situation any pair of observers would be able to predict with certainty what would be the result of the third one, if he also chooses a setting which allows the correlations described above. 
\begin{itemize}\item
For example Anton and Mike measure both vertical  components, and each receives the value $-1$. It they exchange the information about their results, they would know that if the choice of Danny was  to find out the value of the horizontal polarization, his results must be $-1$ (as the product of the three local results must always be $-1$). 
\item However, if Anton and Mike measure vertical and horizontal components, respectively, and they receive, say $-1$ and $1$, they would know (after information exchange)  that if the choice of Danny is to measure the vertical component, he would definitely receive $-1$.
\end{itemize}

{\em Elements of reality. }
This is the moment at which EPR could enter with their definition. They did  say:
"If, without in any way disturbing a
system, we can predict with certainty (i.e., with
probability equal to unity) the value of a physical
quantity, then there exists an element of physical
reality corresponding lo this physical quantity."
Additionally they stressed that "every
element of the physical reality must have a counter
part in the physical theory". Notice that all four situations discussed above are cases in which each result of each observer can be predicted with certainty by other two observers. E.g., in the first displayed example Mike and Anton can fix the value of Danny's element of reality  pertaining to his possible measurement result for horizontal polarization.  Thus this possible result must be "part of the physical theory".  However, the second example fixes the value of the vertical component of spin of Danny's particle. It is also potentially an element of reality.

EPR realized that  ``one  would not arrive
at our conclusion if one insisted that two or more
physical quantities can be regarded as simultaneous
elements of reality only when they can be
simultaneously measured or predicted. On this
point of view, since either one or the other, but
not both simultaneously, of the quantities [horizontal spin component]
and [vertical component] can be predicted, they are not simultaneously
real. This makes the reality of [horizontal component] and [vertical component]
depend upon the process of measurement carried
out on the [two other] system[s], which does, not disturb
the [third] system in any way. No reasonable
definition of reality could be expected to permit
this."\footnote{The phrases in brackets indicate the text of EPR transformed in such a way so that it fits the considered three-particle example. Their $Q$ (position) is now  horizontal component of the spin, and $P$ (momentum) is the vertical one.}

Thus the claim is that any  "reasonable
definition of reality" would treat the values of horizontal and vertical components of the spin as two elements of reality, missing in the quantum theory. Thus quantum theory in incomplete. This where the argument of EPR ends.

Bohr in his 1935 reply said  ``... there is essentially the question of an influence on the very conditions which define the possible types of predictions regarding the future behavior of the system... In fact, it is the mutual exclusion of any two experimental procedures, permitting unambiguous definition of complementary physical quantities, which provides room for new physical laws the coexistence of which at first sight appear irreconcilable with the basic principles of science.''

Thus, the trouble is that in the considered examples, leading to elements or reality,  Mike is to measure either vertical or horizontal components of the spin of his particle. However no experimental device can measure the two components
simultaneously (as such a device must must have the property that it singles our a specific direction, linked with the measured spin component; there is no way to {\em single} out {\em two} directions).  The two components are complementary, they are non-commensurable\footnote{The mathematical formalism of quantum mechanics reflects complementarity of pairs ``observables".  If this is the case say for observables $A$ and $B$, e.g. describing  two different components of spin, then they ``do not commute". This is turn means that in  the formal quantum description ``operators" associated which the two observables we have the following property: $AB\neq BA$. Of course complementarity can occur in various degrees. We have perfect complementarity when experiments measuring B give completely random results for quantum systems prepared in any state, which was prepared by measuring A and selecting only systems which gave the same result of this measurement. For example, photons which are selected by a polarization analyzer which allows only vertically polarized photons to pass through it, would upon subsequent measurement of circular polarization give fully random results. Either clockwise or anti-clockwise polarized photons would appear, with equal probabilities. like in a coin toss.  }.

So, who is right EPR of Bohr? The statement of  EPR is equivalent to treating two situations the actual one (say, Mike in a given run of the experiment, on a specific particle, measured vertical component) and the
 potential one  (Mike  could have measured instead the horizontal component) on equal footing. Replacing ``this" in the quoted paragraph of EPR by what is actually meant by it we get: ``No reasonable
definition of reality could be expected to permit
[...]" "that two or more
physical quantities can be regarded as simultaneous
elements of reality only when they can be
simultaneously measured or predicted." Thus EPR effectively treat the actual and potential different situations [measurements] on equal footing\footnote{Such an approach accepts so called ``counterfactual" statements or conditionals. Such statements contain an "if" clause which describes a situation which in fact did not occur: e.g., "If EPR knew the results of the GHZ paper, they would not have written their 1935 work". }. Bohr definitely does not.

Who is right? Let us return to the GHZ predictions for three spins. As all possible combinations of the results which are consistent with the quantum predictions are equiprobable (recall that values obtained by each observer have the statistics of a fair coin toss), the EPR elements of reality for a specific run of an experiment and the first three potential situations (in which only one of the observers measures the horizontal component, other two measure vertical) can be all $-1$. Why not? Their product is $-1$, everything OK. However, this means that if any one of them measures the horizontal component, and the notion of elements of reality is internally consistent and counter-factual reasonings allowed, then the result must be $-1$ (as {\em all} results mentioned above were assumed to be  such, and one of them in all three situations pertains to measurement of the horizontal component by one of the observers, in each case a different observer). This implies that if by their own whim in the considered run of the experiment   accidentally  all choose to measure horizontal components (the fourth situation), the product of their results must be $-1$. However quantum mechanics predicts the product to be is such a case is {\em always} $+1$, see above. The reader may check all other combination of possible results which agree with the three first situations considered in our example. This invariably leads to the the same prediction for the fourth situation: if elements of reality are to be a consistent notion then  the product must be $-1$. 

Thus if the thesis of EPR holds then $1=-1$. As the complementary nature of horizontal and vertical spin components prohibits the reasoning of the previous paragraph: if Bohr is right  one still has $1=1$. If for a given photon Mike measures the horizontal component, he is not allowed to even speak about the possible value for the complementary measurement. 

This is the end. Where here is non-locality? One could try to argue that non-locality is the solution of this conundrum, provided one insists that counterfactual situations can be treated on equal footing with the actual ones. Then the fact that Mike and Anton choose to measure horizontal component could, by  "a spooky action at a distance",  flip the value of Danny's element of reality for the horizontal component to $+1$.  Of course, such a non-locality would clearly contradict EPR argument as: "This makes the reality of [horizontal component] and [vertical component]
depend upon the process of measurement carried
out on the [two other] system[s], which does, not disturb
the [third] system in any way. No reasonable
definition of reality could be expected to permit
this."

Thus non-locality cannot be derived via an EPR-type reasoning, just as the elements of reality (and thus determinism) are not derivable. The bad luck of EPR was to consider in their work the specific quantum state and specific ``observables", for which elements of reality seem to be a consistent notion (the original EPR state was a state of two  particles with total momentum equal to zero). In 1964 Bell took the simplest possible two spin entangled state, the so-called singlet, and showed that if we accept EPR reasoning and thus elements of reality, then the bound for his original inequality for elements of reality does not hold for quantum predictions (i.e., the inequality is violated). However his reasoning was not as simple as the one for GHZ states, which allows to reveal directly the fallacy of EPR ideas.

The moral of this story is that with EPR-type reasoning it is {\em not } possible to derive determinism, elements of reality, etc. This is so despite the fact the tacit assumption of EPR is counterfactual definiteness, which  is effectively  a rejection of complementarity, as in counterfactual reasonings  we can talk about values of measurements which were actually {\em not} done. So the situation is really tragic. The EPR reasoning, which originally already assumed a form of realism (counterfactual definiteness) when confronted with Bell and GHZ reasonings cannot lead us to a consistent notion of elements of reality, as they imply $1=-1$. All this  invalidates one of their assumptions. If we carefully enumerate the assumptions they are: ``free will", countefactual definiteness, locality and quantum predictions. Their conjunction must be false. As quantum mechanics has not been as yet falsified in any experiment, and ``free will" is rather indisputable, it is locality and/or counterfactual definiteness which must be abandoned. But which - there is no answer.

Additionally, we see the EPR aim to falsify complementarity cannot be reached once we consider GHZ correlations. They wrote "Starting then
with the assumption that the wave function
does give a complete description of the physical
reality, we arrived at the conclusion that two
physical quantities, with noncommuting operators,
can have simultaneous reality." However, their effective assumption of counterfactual definiteness directly implies rejection of complementarity. Thus, the reasoning does not lead to any progress in this question (effectively they show that counterfactual definiteness prohibits complementarity, which is a tautology). Still, with their example, of two particles with vanishing total momentum,  everything seems to be internally consistent, although... circular. However, had they considered the GHZ correlations, they would have not been able to reach the consistency, as in such a case we  are led to $1=-1$, which is a pretty inconsistent statement in elementary algebra.


\section{Local causality}

The above  fairy tale shows that any attempt to derive determinism, or realism via EPR correlation is futile (as just one counterexample is needed to disproof such a claim\footnote{EPR forgot that if a new notion is to be introduced to a theory, then it must checked whether it is consistent with {\em all } predictions of the theory...}). Thus realism of a form  must be separately  assumed in any derivation of Bell inequalities. This means that  violation of such inequalities  is a falsification of a compound assumption which is a conjunction realism of sorts (at least counterfactual definiteness), locality (no action at a distance, constraints of relativistic causality working), and free will (settings decided at the whim of the local observer, or existence of stochastic processes which decide the local  settings, which can be statistically independent of any other factor in the experiment)\footnote{``Free will" is usually not a challenged assumption, thus we assume it to hold throughout the discussion.}.

Still, there is a current fashion to claim that what is sufficient to derive Bell's inequalities is just freedom to choose settings and {\em local causality}, which is treated as unrelated with any form of realism or hidden variables.

Let me therefore present a standard introduction of local causality, which mainly follows the work of Bell `La nouvelle cuisine'~\cite{Bell2004}. 

We have two space-like separated parties, Alice and Bob. They can choose freely   between a number of local  measurement settings. Let us denote, by $x$ and $y$ Alice's and Bob's  measurement settings and by $A$ and $B$ their 
results. The predictions for a Bell type experiment are given probabilities $p(A,B|x,y)$. 

It is argued, that the probabilities $p(A,B|x,y)$ are  a statistical mixture of different situations, labeled by $\lambda$, and  called by Bell `causes'\footnote{{} Note already here, that $\lambda$'s do not appear in quantum mechanics, thus they are {\em hidden variables. Basically this could already end the discussion, as hidden variables are a program of completing quantum mechanics, just like the aim of EPR. As a matter of fact elements of reality are indeed hidden variables. }}.
The probabilities  acquire the following form $$p(A,B|x,y)=\int d\lambda \rho(\lambda) p(A,B|x,y,\lambda),$$ where $\rho(\lambda)$ is a probability distribution of the causes. Standard  formulas for  conditional probabilities, and the fact that conditional probabilities for the same condition have all properties of unconditioned probabilities,  allow one to put $$ p(A,B|x,y,\lambda)= p(A|B,x,y,\lambda) p(B|x,y,\lambda).$$ 
Local causality assumption as stated by Bell reads `The direct causes (and effects) of [the] event are near by, and even the indirect causes (and effects) are no further away than permitted by the velocity of light'. This  allows one to state that `what happens on Alice's side does not depend on what happens on Bob's side' and {\em vice versa}~\cite{Gisin}. Thus the following must hold
$p(A|B, x, y, \lambda) = p(A|x,\lambda)$ and $p(B|x,y,\lambda)=p(B|y,\lambda)$. By symmetry, which must be assumed in any reasonable approach   $p(B|A, x,y,\lambda)=p(B|y,\lambda).$ Thus, we obtain the general mathematical structure of probabilities  which allows to derive all two-particle Bell inequalities:
\begin{equation}
p(A,B|x,y)=\int d\lambda \rho(\lambda) p(A|x,\lambda) p(B|y,\lambda),
\label{condition}
\end{equation}
provided one additionally assumes ``free will" to choose measurement settings.

Sometimes, local causality is thought to be synonymous to locality. There are claims that introduction of this notion by Bell in 1976 is effectively his second theorem about entanglement\footnote{Some authors reserve the phrase Bell's second theorem to his independent derivation of the impossibility of non-contextual hidden variables.}. This I cannot understand because stochastic hidden variable theories, giving probabilities of the structure of Eq. (\ref{condition}), were introduced by Clauser and Horne in 1974 \cite{CH}. The formula $p(A|B, x, y, \lambda) = p(A|x,\lambda)$ is already implied by in the Bell 1964 condition: local result depends on the local setting and the local hidden variables $\lambda$  (or ``more complete specification"), however this was formulated by Bell in the deterministic context (just replace here `local result' by `probability of a local result', and local causality emerges). I shall argue below the local causality is a form of local realism or local hidden variables\footnote{Of course there is a full mathematical equivalence between local causal theories and stochastic local hidden variable theories of Clause and Horne. I shall argue that additionally there is no conceptual difference.}

The danger of in thinking that local causality is equivalent to locality is the fact the opposite notion  to locality is non-locality. However violation of local causality implies either non-local causality, or that we have spontaneous events of local or non-local nature and/or influences.  Local causality assumes locality and existence of causes $\lambda$, which are not present in the quantum formalism.

In the case of {\em mixed} separable states, one may think that $\lambda$ specifies the ``actual'' quantum state in the probabilistic mixture. All this would agree with the formula~(\ref{condition}). However when the two particle quantum state is a pure  entangled one, denoted here by $\psi$, there is a trouble. There is just one joint quantum mechanical state describing the two separated systems. No other specification of the situation is allowed in quantum mechanics, all predictions are derivable using the state and the quantum formalism which additionally gives us methods of calculation probabilities, once the state is known. The probabilities are  calculable using projectors which depend on the (local) settings. Thus the quantum state $\psi$ is the sole ``cause" in quantum theory, except the settings. Applying local causality principle would mean that $p(A|B, x, y, \psi) = p(A|x,\psi)$ and equivalently $p(B|A, x, y, \psi) = p(B|x,\psi)$, and the formula (\ref{condition}) would read
\begin{equation}
p(A,B|x,y)= p(A|x,\psi) p(B|y,\psi),
\label{condition2}
\end{equation}
implying no correlations whatsoever!
To get correlations, one must introduce at least one {\em two valued} `cause', say $\lambda=\lambda_1$ or $\lambda_2$, other than $\psi$. In this way we can get
\begin{equation}
p(A,B|x,y)= \sum_{i=1}^2p(A|x,\psi, \lambda_i) p(B|y,\psi, \lambda_i).
\label{condition2}
\end{equation}
Such a formula allows for correlations.
As such additional $\lambda$'s do not appear in quantum formalism, they are hidden variables {\em per se}. They are an attempt to complete the quantum formalism by some additional factors.

 The $\lambda$'s , which enter Eq.~(\ref{condition}) get various names: e.g. `the physical state of the systems as described by any possible future theory'~\cite{Gisin}, `local beables' \cite{BELL-BEABLES}, `the real state of affairs', `complete description of the state', etc.
 Bell himself writes `$\lambda $ denote any number of hypothetical additional  variables needed to complete quantum mechanics in the way envisaged by EPR'~(\cite{Bell2004}, page 242). This sentence of Bell's is often forgotten by those who think that local causality differs from local hidden variable theories. As a matter of fact even Bell himself had a tendency to ignore it \cite{ZUK-SHPP}.

The other way of looking at this is to notice that local causality implies existence of an underlying joint probability distribution for results of all possible measurements (commensurable or non-commensurable), which is normalizable to unity and non-negative. Once we have probabilities $ p(A|x,\lambda)$ and $ p(B|y,\lambda)$  to define such an object is trivial, 
while in general in quantum mechanics this in general  is impossible, and there is no quantum mechanical method to formulate such a distribution.
Denote by $A_{x_i}$ the possible values (spectrum) of an observable $\hat{O}^A_{x_i}$ for Alice's system, and also   by $B_{y_j}$ the possible values (spectrum) of an observable $\hat{O}^B_{y_i}$ for Bob's system. Indices $i,j$ numerate the observables in whatever way. 
Then the underlying distribution reads

$$P(A_{x_1},..., A_{x_n},B_{y_1},..., B_{y_m} )=\int d\lambda \rho(\lambda) \prod_{i=1}^{n}  p(A_{x_i}|x_i,\lambda)\prod_{j=1}^{m}  p(B_{y_j}|y_j,\lambda)$$
and all other probabilities in a local causal theory are marginals of such distributions. If in a probability theory such distributions exist then then all properties of the axioms of Kolmogorov are satisfied, and probabilities can have the classical lack of knowledge interpretation. One can model them by a normalized measure 
on some sample space $\Omega$. Elements of $\Omega$ have all properties of hidden variables. Thus conversely any theory with all  $P(A_{x_1},..., A_{x_n},B_{y_1},..., B_{y_m} )$ existing and non-negative can be modelled with some hidden variable on a sample space, for details see \cite{CASLAV-MZ}.

\section{Final remarks}
In Section 2 it was shown that the EPR reasoning cannot be used to derive determinism or realism of sorts, as it has as a tacit assumption acceptance of courterfactual statements as valid. Once one accepts such statements, then for a given quantum system unmeasured quantities have the same status as the one actually measured, even if they are mutually non-commensurable. This is of course prohibited by complementarity principle. Thus, EPR reasoning cannot be used do deny complementarity, because EPR deny complementarity in their initial assumptions.

Still the situation is even worse, as in the counterfactual  situation of EPR considering the GHZ correlations, they would have run into a $1=-1$ contradiction in their denial of complementarity. Thus, EPR reasoning, as it leads in this case to an outright contradiction cannot be useful in any scientific reasoning, and especially it is useless as a tool to derive determinism or realism of any sort (as this leads in such a case to a contradiction). Still, the works is seminal and extremely important - but these are not terms which describe its internal consistency, especially  if one considers all quantum predictions, and not only for their state (in the case of which unluckily for the history of physics the contradiction was difficult to spot\footnote{For a version Bell's  theorem for EPR states see \cite{BANASZEK}, or for a more recent development see \cite{ROSOLEK}.}).

Thus, the consequences of the inconsistency of EPR argument are either that  unperformed experiments have no results \cite{UNPERFORM}, or non-locality, or both (if we insist to retain ``freedom"). With counterfactual reasonings the algebraic identity leading to the CHSH \cite{CHSH} inequality reads:
\begin{equation}
A_{x_1} B_{y_1}+A_{x_1} B_{y_2}+A_{x_2} B_{y_1}-A_{x_2} B_{y_2}=\pm2.
\end{equation} 
Note that it pertains to all possible choice of local measurements  in two-setting per observer Bell experiment, hypothetically applied in the case of a given single pair  of particles.  With complementarity in mind, and for the case in which the actual setting for the given pair of particles is on both sides the first one, $x_1, y_1$, as  values for the complementary settings are then {\em unspeakable}, we have
\begin{equation}
A_{x_1} B_{y_1}+A_{x_1} (?)+(?) B_{y_1}-(?)(?)=?.
\end{equation} 
.

In the considerations of Section 3, it is shown  that local causality is compound notion. Thus,  its opposite is not `non-locality'. Local causality is equivalent to stochastic local hidden variable theories introduced in \cite{CH}. We have locality and causes  $\lambda$ which are not present in quantum description, and therefore are hidden variables. Thus an antonym of local causality may be  non-locality or spontaneous  events, or both.

The author is supported by FNP (project TEAM), and  ERC AdG grant QOLAPS.

\end{document}